\documentclass[noeprint,twocolumn,prb,showpacs,preprintnumbers,superscriptaddress,amsmath,amssymb,10pt,aps]{revtex4-1}
\usepackage{graphicx}
\usepackage{dcolumn}
\usepackage{color}
\usepackage{bm}
\usepackage[hidelinks]{hyperref}
\hypersetup{
    colorlinks,
    citecolor=blue,
    filecolor=blue,
    linkcolor=blue,
    urlcolor=blue
}

\newcommand{\ud}{\,\mathrm{d}}
\newcommand{\HH}{\mathcal{H}}
\newcommand{\N}{\mathcal{N}}
\newcommand{\e}{\textrm{e}}
\newcommand{\im}{\textrm{i}}

\newcommand{\CC}{\mathcal{C}}
\newcommand{\PP}{\mathcal{P}}
\newcommand{\TT}{\mathcal{T}}
\newcommand{\KK}{\hat{K}}

\DeclareMathOperator{\hside}{\Theta}

\DeclareMathOperator{\sgn}{sgn}

\begin{document}

\title{Chiral Hall effect in the kink states in topological insulators with magnetic domain walls}

\author{M. Sedlmayr}
\email{m.sedlmayr@prz.edu.pl}
\affiliation{Department of Physics and Medical Engineering, Rzesz\'ow University of Technology, al.~Powsta\'nc\'ow Warszawy 6, 35-959 Rzesz\'ow, Poland}
\author{N. Sedlmayr}
\affiliation{Institute of Physics, M.~Curie-Sk{\l}odowska University, 20-031 Lublin, Poland}
\author{J. Barna\'s}
\affiliation{Faculty of Physics, Adam Mickiewicz University, ul. Uniwersytetu Pozna\'nskiego 2, 61-614 Pozna\'n, Poland}
\author{V. K. Dugaev}
\affiliation{Department of Physics and Medical Engineering, Rzesz\'ow University of Technology, al.~Powsta\'nc\'ow Warszawy 6, 35-959 Rzesz\'ow, Poland}

\begin{abstract}
In this article we consider the chiral Hall effect due to topologically protected kink states formed in topological insulators at boundaries between domains with differing topological invariants. Such systems include the surfaces of three dimensional topological insulators magnetically doped or in proximity with ferromagnets, as well as  certain two dimensional topological insulators. We analyze the equilibrium charge current along the domain wall and show that it is equal to the sum of counter-propagating equilibrium currents flowing along external boundaries of the domains.  In addition, we also calculate the current along the domain wall when an external voltage is applied perpendicularly to the wall.
\end{abstract}

\date{\today}

\maketitle

\emph{Introduction.} In recent years topological insulators and superconductors have attracted a considerable amount of attention, \cite{Hasan2010,Fu2006,Fu2007,Qi2011} driven by both interest in their fundamental properties, and by potential applications for quantum computing and spintronics. The most visible consequence of a topological insulator being in a topologically non-trivial phase is the appearance of protected gapless edge states. For a three-dimensional topological insulator these often manifest themselves as two-dimensional Dirac cones for the electrons confined to the surface.\cite{Zhang2009} Magnetic impurities can locally gap the surface states,\cite{Liu2009} while magnetic doping at a moderate level can open the gap fully leading to massive Dirac fermions on the surface\cite{Chen2010a} and interesting interface properties.\cite{Henk2012,Rauch2013} Furthermore, magnetic impurities can be either ferromagnetically ordered or disordered.\cite{Abanin2011,Cheianov2012,Rosenberg2012,Zhang2012a,Assaf2015} In turn, transport properties of ferromagnet-topological insulator layers  reveal large magnetoresistance effects,\cite{Kong2011,Zhou2014,Rzeszutko2017} which are interesting from the application point of view. Of some interest are also superconducting proximity effects which can be very long ranged in the topological surface states.\cite{Dayton2016}

We consider a heterostructure consisting of a three-dimensional topological insulator and a ferromagnetic layer\cite{Yasuda2017a,Litvinov2020} with magnetic domains present. One can find topologically protected states confined to the lines following the magnetic domain boundaries of the ferromagnet. These states have interesting properties and may be relevant for the construction of novel spintronic devices.\cite{Yasuda2017} It has been shown that the presence of a topological insulator beneath a thin ferromagnetic film increases the Walker breakdown threshold for the motion of the domain walls in the ferromagnet.\cite{Linder2014,Ferreiros2014,Ferreiros2015} This, in turn, allows for increased domain wall velocities. Magnetisation dynamics and switching of ferromagnetic thin films on topological insulators has also received a lot of attention.\cite{Yokoyama2011,Tserkovnyak2012,Fan2014,Semenov2014} Similarly, heterostructures involving topological insulators, heavy metals, and ferromagnets are also investigated for their spin torque properties and potential applications.\cite{Mellnik2014,Mahfouzi2016}

\begin{figure}
\includegraphics[width=0.95\columnwidth]{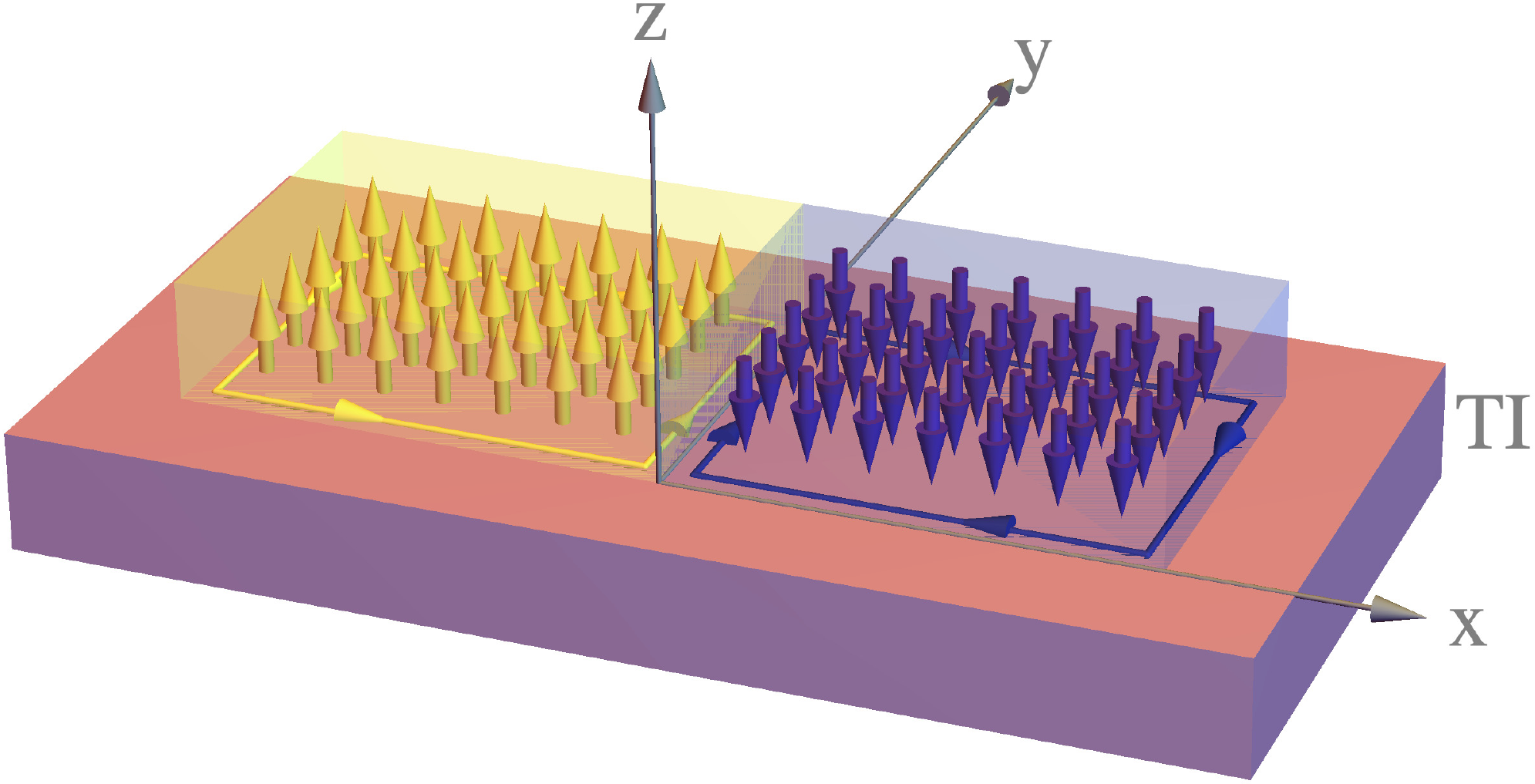}
\caption{Schematic of a topological insulator-ferromagnet heterostructure. Two ferromagnetic domains with opposite magnetisation direction are placed on top of a topological insulator. Edge currents along the domain wall and along external boundaries of the domains are shown in yellow and purple for the two ferromagnets. The topologically protected kink states form on the surface of the topological insulator along the domain wall boundary.}
\label{schematic}
\end{figure}

In this paper, we consider the equilibrium current flowing along the domain wall. This current is a sum of two counter-propagating currents along the edges of the two magnetic regions with opposite magnetisations, see Fig.~\ref{schematic}. In addition, we also consider the nonequilibrium current along the domain wall when a voltage is applied across  the domain wall (perpendicularly to the wall), and calculate  the anomalous Hall conductance for the kink states confined to the domain wall. We obtained good agreement with available experimental results. The anomalous Hall effect (AHE) exits in the absence of external magnetic field, and appears due to internal magnetisation.\cite{Yu2010,Nagaosa2010,Xue2013,Checkelsky2014,Kou2014,Chang2015,Bestwick2015,Ou2018} Such effect can also be found in two dimensional hexagonal lattices\cite{Taillefumier2008,Taillefumier2011}, closely related to the models we consider here. Recent experiments have reported large AHE in topological insulators with proximity induced magnetism.\cite{Yasuda2017a,Mogi2019,Yao2019a} In the model studied here the AHE appears due to a chiral magnetic structure of the domain wall. Therefore, to distinguish is from the usual AHE, we call it the  chiral  Hall effect (CHE).

\emph{Model.} The two-dimensional surface states of a three-dimensional topological insulator, for example those of Bi$_2$Se$_3$ or Bi$_2$Te$_3$, can be described, by making a unitary transformation to an appropriate spin basis, by a simple effective theory:\cite{Zhang2009,Wakatsuki2015} $\HH_{\rm TI}=v\mathbf{k}\cdot{\bm\sigma}$, where ${\bm\sigma}=(\sigma^x,\sigma^y,\sigma^z)$ are the Pauli spin matrices and $\mathbf{k}=(k_x,k_y,0)$ is the electron wavevector.  The rescaled velocity is $v=\hbar v_F$ with $v_F$ the Fermi velocity of the surface states. For Bi$_2$Se$_3$ this is $v_F=3.6\times10^5$ ms$^{-1}$.\cite{Zhang2009,Zhang2011c} By placing a ferromagnet on top of the three-dimensional topological insulator, see Fig.~\ref{schematic}, the stray field from the ferromagnet introduces a local Zeeman field, $\mathbf{m}(x,y)=gM(x,y)\hat z$, that is determined by the $z$-component of magnetization $M_z(x)$ (here $g$ is the coupling constant), which results in
\begin{equation}\label{tifm}
	\HH_{\rm TI-FM}=v \mathbf{\hat k}\cdot{\bm\sigma}+\mathbf{m}(x,y)\cdot{\bm\sigma}\,,
\end{equation}
with $\mathbf{\hat k}=-\im(\partial_x,\partial_y,0)$.  The stray field follows the structure of the magnetic domains in the ferromagnet, though here we will focus on the simple situation,
\begin{equation}
	\mathbf{m}(x,y)=\hat{\bf z}m(x)=\hat{\bf z}\left\{ \begin{array}{cc} m_0\,, & x<0\,, \\ -m_0\,, & x\ge 0 \,,\end{array}  \right.
\end{equation}
valid for a continuous rotation of magnetization at the domain wall, provided that the domain wall width is shorter than the localisation length-scale of the kink states. $\hat{\bf z}$ is a unit vector along the $z$-axis.

Additionally, we assume an electronic potential localized at the domain wall: $V(x)= V_0$ if $|x|<\delta $ and $V(x)=0$ if $|x|\ge \delta$, where $2\delta $ is the barrier thickness. Assuming $\kappa \delta \ll 1$ ($\kappa$ is an attenuation factor to be defined below) one can consider $V(x)$ as a $\delta$ potential, and in this limit one may write $V(x)\to 2v\lambda\delta(x)$, where $\lambda=\frac{V_0}{v}\delta$. Finally, we assume  a voltage applied across the wall, $\mu(x)=-\delta\mu$  if $x<-\delta$ and $\mu(x)=\delta\mu$ for $x > \delta$, where $\mu$ is the relevant electrochemical potential. Thus, the final form of the Hamiltonian is
\begin{equation}\label{leham}
	\hat{\HH}=-\im v\left( {\sigma}^x \partial_x+{\sigma}^y\partial_y\right)+m(x){\sigma}^z+2v\lambda\delta(x)-\mu(x)\,.
\end{equation}
The Schr\"odinger equation for the spinor components of the wavefunction $\psi ^T=\big(\varphi,\, \chi\big)$ is  $(\hat{\HH}-\varepsilon )\, \psi ({\bf r})=0$. Due to the translational invariance along the $y$-axis we can write $\psi ({\bf r})=e^{ik_yy}\psi_{k_y}(x)$.

Let us now turn to the symmetries of the effective model \eqref{leham}, and consider the homogeneous bulk form of  Eq.\eqref{leham}:
\begin{equation}\label{lehambulk}
	\hat{\HH}=v\left(k_x{\sigma}^x+k_y{\sigma}^y\right)+m{\sigma}^z-\mu\,.
\end{equation}
For $m=\mu=0$ we have time reversal symmetry $[\TT,H]=0$ with $\TT=\im{\sigma}^y\KK$, which is broken for $m\neq0$. This should be clear for the heterostructure as in that case $m$ is caused by the stray field of a ferromagnet. There is a particle-hole symmetry $[\CC,\HH]_+=0$ for $\CC={\sigma}^x\KK$, which is broken by the presence of a non-zero chemical potential $\mu$. Finally we still have chiral symmetry $[\PP,\HH]_+=0$ with $\PP=\CC\TT={\sigma}^z$. Naturally this is broken for either non-zero $m$ or $\mu$.

To find edge currents (equilibrium and nonequilibrium ones) we need to first find the electronic edge states, and first we consider the kink states bound to the domain wall. The solutions on either side of the barrier are
\begin{equation}\label{ansatz}
	\psi_{k_y}(x>,<0)=\frac{A_\pm \e^{-\kappa_\pm |x|}}{\sqrt{2}}
	\left(\begin{array}{c} \varepsilon_\pm\mp m_0 \\ \pm\im v(\kappa_\pm\pm k_y) \end{array}\right)\,,
\end{equation}
where
$v\kappa_\pm=(m_0^2-\varepsilon_\pm^2+v^2k_y^2)^{1/2}$, $\varepsilon_\pm=\varepsilon\pm\delta\mu$, and $A_\pm$ are parameters to be determined from the continuity conditions. To match the solutions for $x<-\delta$ and $x>\delta$, we integrate the Schr\"odinger equation with Hamiltonian \eqref{leham} on $x$ from $x=-\delta $ to $x=\delta$, and then take the limit $\delta\to0$. Then we find for $\psi=(\varphi,\chi)^T$:
\begin{eqnarray}
	\varphi (\delta )-\varphi (-\delta )&=&-\im\frac{\lambda}{v}\chi (0)\, \\
	\chi (\delta )-\chi (-\delta )&=&-\im\frac{\lambda}{v}\varphi (0)\,.
\end{eqnarray}
One can take $\varphi (0)=\frac12 [\varphi (\delta )+\varphi (-\delta )]$ and $\chi (0)=\frac12 [\chi (\delta )+\chi (-\delta )]$. Solving the boundary conditions leads to the following dispersion relation for the topologically protected kink states (for $m_0>0$),
\begin{equation}\label{dispersion}
	\varepsilon=\frac{vk_y(1-\lambda^2)-2m_0\lambda}{1+\lambda^2}\sqrt{1-\frac{\delta\mu^2(1+\lambda^2)^2}{\left[m_0(1-\lambda^2)+2\lambda v k_y\right]^2}}\,,
\end{equation}
and to the corresponding normalized wave functions,
\begin{equation}\label{final_wfn}
	\psi_{k_y}(x>,<0)=\frac{\e^{-\kappa_\pm |x|}}{\sqrt{2}\N}
	\frac{1}{\lambda+\gamma_\pm}
	\left(\begin{array}{c} 1 \\ \pm\im \gamma_\pm \end{array}\right)\,,
\end{equation}
where $\gamma_\pm=v \kappa_\pm\pm k_y/(\varepsilon\pm\delta\mu\mp m_0)$, and
\begin{equation} \N=\frac{1}{2}\sqrt{\frac{1+\gamma_+^2}{\kappa_+\left(\lambda+\gamma_+\right)^2}+\frac{1+\gamma_-^2}{\kappa_-\left(\lambda+\gamma_-\right)^2}}\,.
\end{equation}

The dispersion relation \eqref{dispersion} simplifies in several limits. For $\delta\mu=\lambda=0$ we have a trivial linear dispersion relation: $\varepsilon=vk_y-\mu$. As we can see there is a symmetry $\varepsilon (k_y)=-\varepsilon (-k_y)$ but $\varepsilon (k_y)\ne \varepsilon (-k_y)$. If we include a potential barrier at the boundary,  but take $\delta\mu=0$, then
\begin{equation}
\varepsilon=\frac{vk_y(1-\lambda^2)-2m_0\lambda}{1+\lambda^2}\,.
\end{equation}
In turn, if we assume a nonzero $\delta\mu$, but take $\lambda=0$, then the dispersion relation acquires the form
\begin{equation}
\varepsilon=vk_y\sqrt{1-\frac{\delta\mu^2}{m_0^2}}\,.
\end{equation}

Since our intention is to demonstrate that the equilibrium current (for $\delta\mu =0$) along the domain wall is formed from counter-propagating equilibrium currents at the edges of the two ferromagnetic regions of opposite magnetisation, see Fig.~\ref{schematic},  we need to consider now the electronic states at the boundary between the region underneath the ferromagnet and where no ferromagnetic layer is present. For this we use the Hamiltonian
\begin{equation}\label{leham1}
	\hat{\HH}_{\rm E}=-\im v\left( {\sigma}^x \partial_x+{\sigma}^y\partial_y\right)+m_0\hside(x) {\sigma}^z\,,
\end{equation}
where $\hside(x)$ is the Heaviside theta function.

The Hamiltonian \eqref{leham1} has eigenstates $\hat{\HH}_{\rm E}\psi^{\rm E}=\epsilon\psi^{\rm E}$, and for $x<0$ we find
\begin{equation}
	\psi^{\rm E}_{k_x,k_y}(x)=a\e^{\im k_xx}\begin{pmatrix}
		1\\ \frac{\epsilon}{v\left(k_x-\im k_y\right)}
		\end{pmatrix}
		+b\e^{-\im k_xx}\begin{pmatrix}
		1\\ \frac{-\epsilon}{v\left(k_x+\im k_y\right)}
	\end{pmatrix}\,,
\end{equation}
while  for $x>0$
\begin{equation}
	\psi^{\rm E}_{k_x,k_y}(x)=c\e^{-\kappa x}\begin{pmatrix}
		1\\ -\frac{\im(\epsilon-m_0)}{v\left(\kappa-k_y\right)}
		\end{pmatrix}\,.
\end{equation}
The parameters $a$, $b$ and $c$  can be determined from the continuity conditions of the wavefunctions at the boundary $x=0$. We note that $|a|^2=|b|^2$, which is the condition for perfect reflection, as expected. The corresponding eigenenergies are
\begin{equation}
	\epsilon=\pm v\sqrt{k_x^2+k_y^2}\,,
\end{equation}
and from conservation of energy we have the relation $v\kappa=\sqrt{m_0^2-v^2k_x^2}$. Note that translational invariance along $y$ prevents the mixing of different $k_y$ channels.

\emph{Chiral Hall effect.} Having wavefunctions for the kink states localized at the domain wall and those at boundaries of the ferromagnet, we can calculate the currents. Let us analyze first the equilibrium current, and start from the contribution  $j^{{\rm E}y}_{k_x,k_y}$ to the equilibrium current along the boundary  using the current operator $\hat j^y=-ev_F{\sigma}^y$:
\begin{equation}
	j^{{\rm E}y}_{k_x,k_y}=-ev_F\int_0^\infty\ud x{\psi^{\rm E}}^\dagger{\sigma}^y\psi^{\rm E}\,.
\end{equation}
All contributions from $x<0$ integrate to zero. We will focus on the negative energy band only assuming the system has its Fermi energy at zero. This leads to
\begin{equation}
	j^{{\rm E}y}_{k_x,k_y}=\frac{e}{\hbar}|c|^2 v^2\frac{m_0+v\sqrt{k_x^2+k_y^2}}{vk_y\sqrt{m_0^2-v^2k_x^2}-m_0^2+v^2k_x^2}\,.
\end{equation}
As $|m_0|>|\varepsilon|$, we see that $\sgn(j^y_{k_x,k_y})=\sgn(m_0)\sgn(k_y)$. However, since this is not an odd function of $k_y$, there is an equilibrium current which is proportional to $\sgn(m_0)$ as required. As there is only one state for every $k_x$ and $k_y$, we can take $|a|^2=1$, and just the negative $\epsilon$ states.

The total equilibrium current can be found from
\begin{equation}
	J^{{\rm E}y}=\int\frac{\ud k_x}{2\pi}\frac{\ud k_y}{2\pi}j^{{\rm E}y}_{k_x,k_y}\,.
\end{equation}
Note that the $k_x$ integral must be confined to $0<vk_x<m_0$ and otherwise the limits are given by $0<v\sqrt{k_x^2+k_y^2}<\Delta$. $\Delta$ is introduced as a cut-off for the edge states. Performing the integral gives
\begin{equation}
	J^{{\rm E}y}=-\frac{e}{h}\frac{\Delta^2}{8\pi m_0}\left[\frac{|m_0|}{\Delta}\sqrt{1-\frac{m_0^2}{\Delta^2}}-\cos^{-1}\frac{|m_0|}{\Delta}+\frac{\pi}{2}\right]
\end{equation}
for the equilibrium current, which in the limit $\Delta>>m_0$ reduces to
\begin{equation}
	J^{{\rm E}y}\sim-\frac{e}{h}\frac{\Delta}{8\pi}\sgn(m_0)\,.
\end{equation}

The topological insulator-ferromagnet heterostructure under consideration has a Chern number $-\sgn m_0$, which can be related directly to the anomalous Hall conductivity.\cite{Taillefumier2008,Taillefumier2011} In the model here rather than the topologically protected states on the edges of the quantum anomalous Hall insulator, we have kink states between two quantum anomalous Hall insulators with opposite Chern numbers. Bearing in mind that we also intend to calculate nonequilibrium chiral Hall effect, we derive here general formula for current {\it via} the kink states, \eqref{final_wfn}, which includes generally both equilibrium and nonequilibrium terms. Using the current operator along the domain wall, $\hat j^y=-ev_F{\sigma}^y$, we find the current due to a state with momentum $k_y$ as
\begin{equation}
	j^y_{k_y}=-ev_F\int_{-\infty}^\infty\ud x \psi_{k_y}^\dagger{\sigma}^y\psi_{k_y}\,.
\end{equation}
Performing the integral we find
\begin{equation}
j^y_{k_y}=-\frac{ev_F}{2\N^2}\left[\frac{\gamma_+}{\kappa_+\left(\lambda+\gamma_+\right)^2}-\frac{\gamma_-}{\kappa_-\left(\lambda+\gamma_-\right)^2}\right]\,.
\end{equation}
The total current is therefore
\begin{equation}
	J^y=\int_{-\frac{\Delta}{v}}^{k_\mu}\frac{\ud k_y}{2\pi}j^y_{k_y}\,,
\end{equation}
where $\varepsilon_{k_\mu}=\mu$ and the lower integral limit is added as a cut-off of the order of magnitude where the kink-states enter the bulk, for the numerical calculations we use $\Delta=m_0$.

In the limit of $\Delta>>m_0$ we can find the equilibrium current (for $\delta\mu =0$) along the domain wall,
\begin{equation}
	J^y\sim\frac{e}{h}\frac{\Delta}{4\pi}\sgn(m_0)\,,
\end{equation}
for $\lambda=0$. This is exactly twice the current along the edges, in agreement with the conjecture that the current along the domain wall is formed from the currents along the edges.

Now, we consider transport due to a voltage applied across the domain wall. Such a voltage generates not only the current flowing along normal to the wall, but also current flowing along the wall. The latter is of particular interest as it involves topologically protected kink states at the wall and reveals anomalous Hall transport properties. Therefore, further analysis will be limited to this current only. The linear response current can be calculated from the formula
\begin{equation}
	J^y_U=U\int_{-\frac{m_0}{v}}^{k_\mu}\frac{\ud k_y}{2\pi}\left.\frac{\partial j^y_{k_y}}{\partial U}\right|_{U=0}\,,
\end{equation}
where $U=\frac{2\delta\mu}{e}$ is the voltage  drop across the domain wall. However, instead of using the above formula, we will calculate $J^y$ numerically from the exact expressions, which also includes nonlinear range.

\begin{figure}
\includegraphics*[width=\columnwidth]{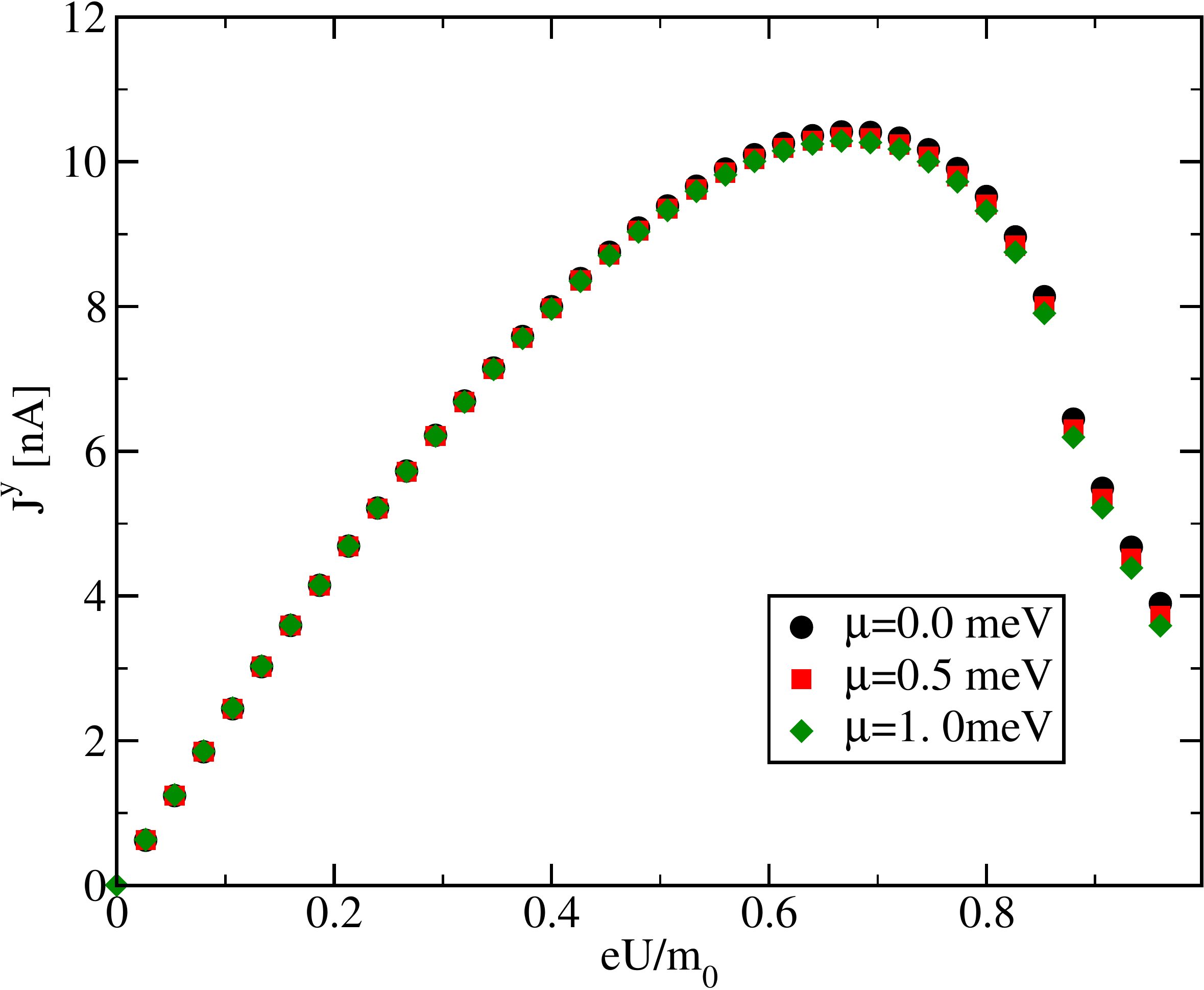}
\caption{Non-equilibrium current along the domain wall $J^y$ as a function of the normalized bias $eU/m_0$, calculated with $v_F=3.6\cdot10^5$ ms$^{-1}$, $m_0=13.5$ meV, and $\lambda=0$. Increasing bias  increases the current until a maximum is reached, the subsequent drop in current is caused by a suppression of the wavefunction on one side of the barrier.}
\label{current}
\end{figure}

\begin{figure}
\includegraphics*[width=\columnwidth]{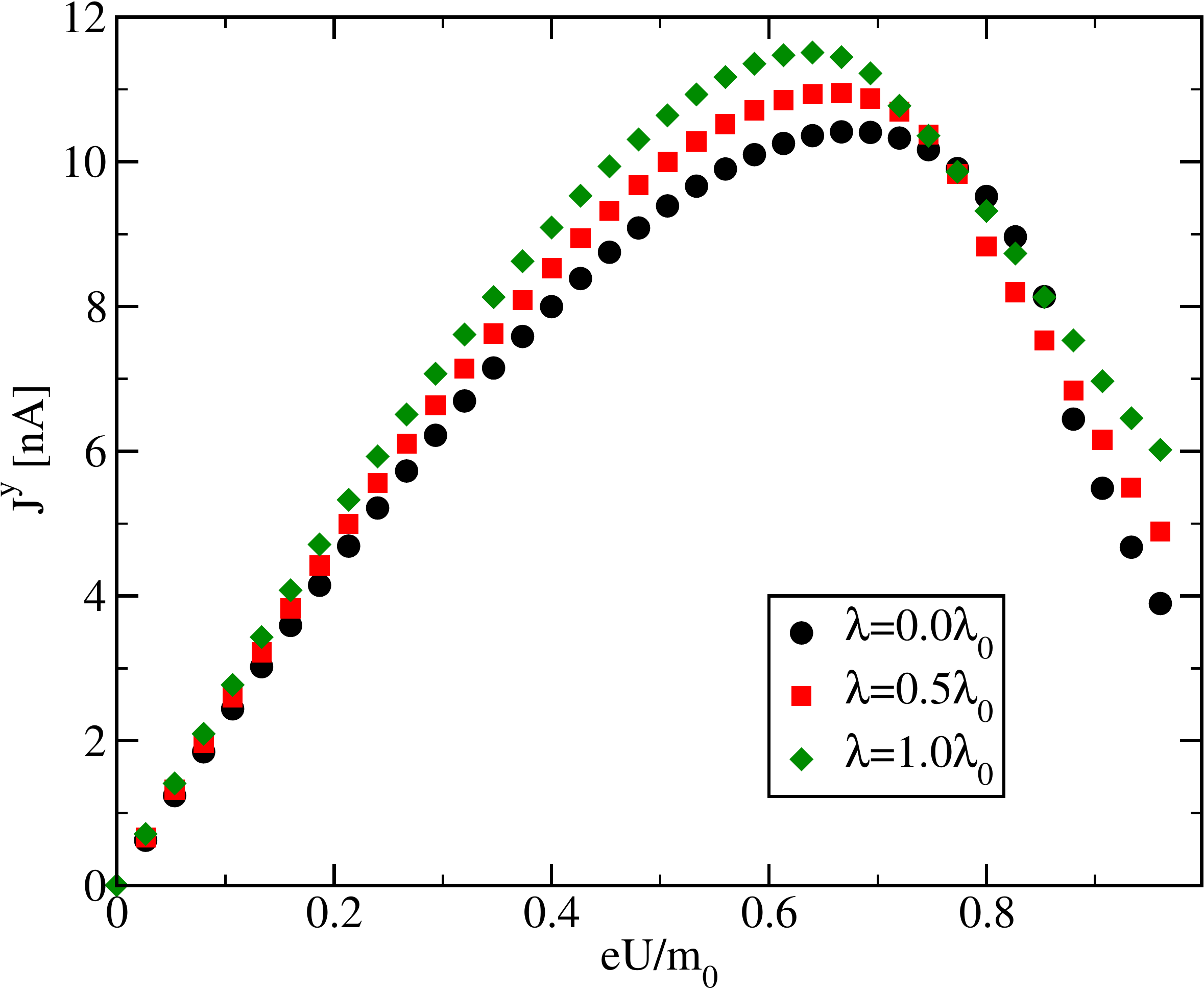}
\caption{Non-equilibrium current along the domain wall $J^y$ as a function of the normalized bias $eU/m_0$, calculated for $v_F=3.6\cdot10^5$ ms$^{-1}$, $m_0=13.5$ meV, $\mu=0$, and $\lambda_0=\frac{V_0\delta}{v}$ with $V_0=1$ meV and $\delta=1$ nm. }
\label{currentlambda}
\end{figure}

In Figs \ref{current} and \ref{currentlambda} we present $J^y$ for several cases as a function of the voltage drop $U$ at the domain wall. We take the Fermi velocity as approximately $v_F=3.6\cdot10^5$ ms$^{-1}$ from Ref.~\onlinecite{Zhang2011c} and the gap induced in the topological insulator surface states by the ferromagnets as $m_0=13.5$ meV from Ref.~\onlinecite{Mogi2019}. The fall in the current when $U$ exceeds a certain value  is caused by a suppression of the wavefunction on one side of the barrier by increasing $\delta\mu$. For the barrier in Fig.~\ref{currentlambda} we used $\lambda_0=\frac{V_0\delta}{v}$ with $V_0=1$ meV and $\delta=1$ nm. Fig.~\ref{current} also shows the optimal range of the voltages for the current.

\begin{figure}
\includegraphics*[width=\columnwidth]{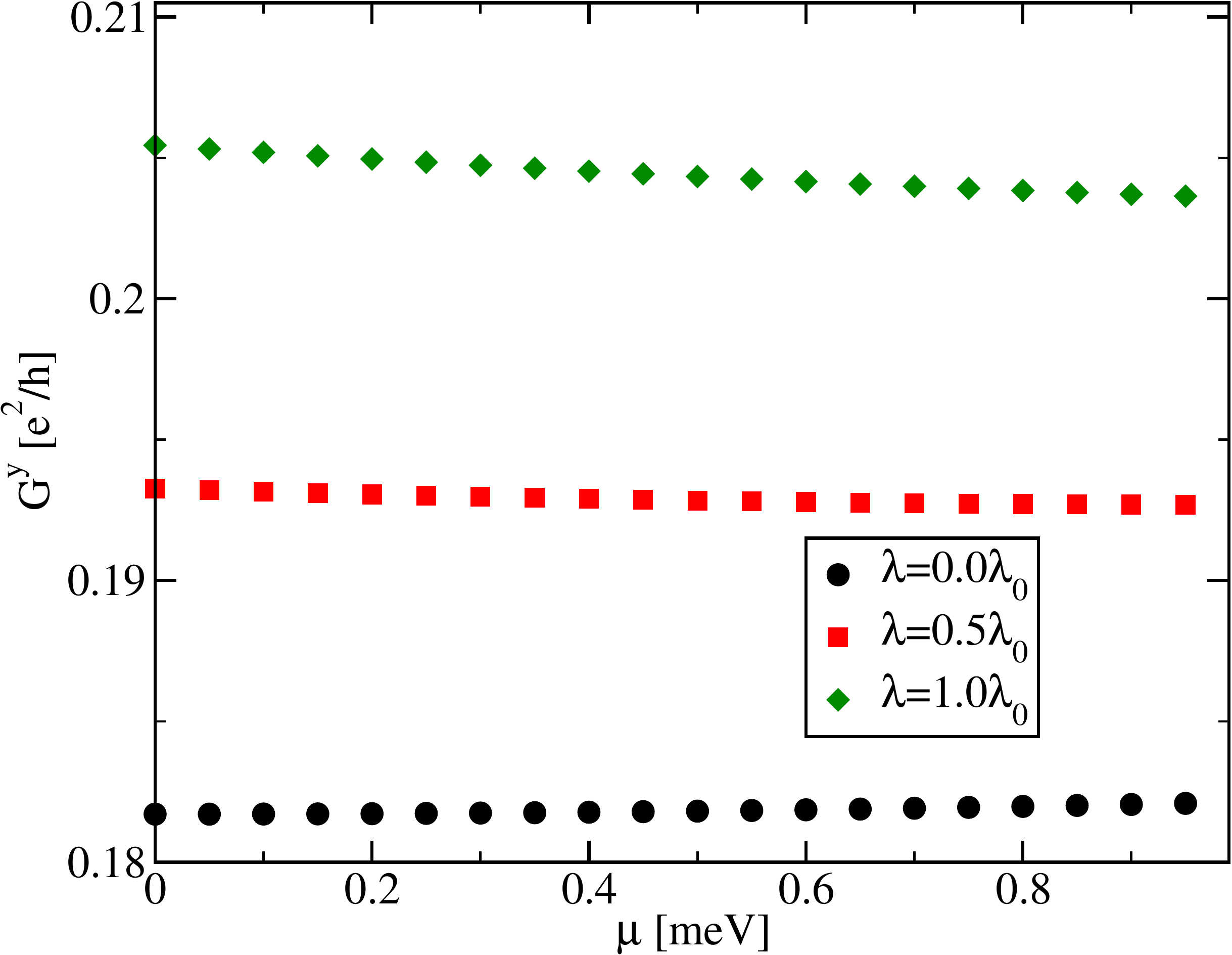}
\caption{Linear chiral  Hall conductance $G^y$  as a function of chemical potential $\mu$ for $v_F=3.6\cdot10^5$ ms$^{-1}$, $m_0=13.5$ meV, and $\lambda_0=\frac{V_0\delta}{v}$ with $V_0=1$ meV and $\delta=1$ nm.}
\label{diffcond}
\end{figure}

From Figs~\ref{current} and \ref{currentlambda} one can extract the CHE conductance $G^y$ in the linear response regime, and we find consistent values of $G^y=0.18\to0.20$ e$^2$/h, similar to those measured in Ref.~\onlinecite{Mogi2019}.  Results for $G^y$ in the linear response range are summarized in Fig.~\ref{diffcond}. As expected changing the chemical potential, and hence the filling of the kink states, does not affect much the conductance. However, a potential barrier at the domain wall makes it easier to form bound states increasing the density of kink states, and hence increasing the conductance, see Fig.~\ref{diffcond}. Differential conductance  $G^y$ in the whole bias range can be determined from Figs~\ref{current} and \ref{currentlambda}  directly as $G^y=\frac{\partial J^y}{\partial U}$.

\emph{Conclusions.} In conclusion, we have calculated the equilibrium current in a three-dimensional topological insulator, flowing along the domain wall induced by a ferromagnet placed on top. This current flows through the topologically protected kink states at the wall, and is shown to be a sum of counter-propagating equilibrium currents flowing along external edges of the two ferromagnetic domains with opposite magnetisations. This current is non-dissipative and may lead to an orbital magnetization, which is measurable. When a voltage is applied across the barrier, not only does a dissipative current flow across the barrier, but also a non-dissipative current flows along the barrier. The latter is a signature of the chiral Hall effect associated with the topologically protected kink states. The calculated conductance is in agreement with available experimental data.

\section{Acknowledgements}
This work was supported in Poland by the National Science Centre under the Project No.~UMO-2017/27/B/ST3/02881. We thank Ingrid Mertig for a critical reading of the paper and insightful comments.

\end{document}